\documentclass[aps,preprint,epsfig,rotate,superscriptaddress,showpacs]{revtex4}
\usepackage{epsfig}
\usepackage{graphicx}
\usepackage{amsmath}

\newcommand{\beq}{\begin{eqnarray}}
\newcommand{\eeq}{\end{eqnarray}}
\begin{document}

\title{Conductance properties of rough quantum wires with colored surface disorder}
\author{Gursoy B. Akguc} \affiliation{Department
of Physics and Centre for Computational Science and Engineering, \\
National University of Singapore, 117542, Republic of Singapore}
\author{Jiangbin Gong} \email{phygj@nus.edu.sg}
\affiliation{Department
of Physics and Centre for Computational Science and Engineering, \\
National University of Singapore, 117542, Republic of Singapore}
\affiliation{NUS Graduate School for Integrative Sciences and
Engineering, National University of Singapore,
 117597, Republic of Singapore}

\begin{abstract}
Effects of correlated disorder on wave localization have attracted
considerable interest. Motivated by the importance of studies of
quantum transport in rough nanowires, here we examine how colored
surface roughness impacts the conductance of two-dimensional quantum
waveguides, using direct scattering calculations based on the
reaction matrix approach. The computational results are analyzed in
connection with a theoretical relation between the localization
length and the structure factor of correlated disorder.  We also
examine and discuss several cases that have not been treated
theoretically or are beyond the validity regime of available
theories. Results indicate that conductance properties of quantum
wires are controllable via colored surface disorder.

\end{abstract}

\date{\today}
\maketitle
\section{Introduction}
Ever since Anderson's model of electron transport in disordered
crystals \cite{Anderson}, wave localization in disordered media has
attracted great interests due to their universality. For example,
two recent experiments directly observed matter-wave localization in
disordered optical potentials using Bose-Einstein condensates
\cite{BECexp1,BECexp2}. One of the most known results from
Anderson's model is that in one-dimensional (1D) disordered systems,
the electron wavefunction is always exponentially localized and
hence does not contribute to conductance for any given strength of
disorder. Note however, this seminal result is based on the strong
assumption that the disorder is of the white noise type. If the
disorder is colored due to long-range correlations, then a mobility
edge may occur in one-dimensional systems as well \cite{izprl}.

Quantum transport in nanowires is of great interest due to their
potential applications in nanotechnology. In addition to the
possibility of ballistic electron transport, quantum nanowires are
found to show many other important properties. In particular,
silicon nanowires  can have better electronic response time
\cite{Ramayya} as well as desirable thermoelectric properties
\cite{thermo}.  It is hence important to ask how the nature of
surface disorder of quantum wires, modeled by quantum waveguides in
this study,  affects their conductance properties.

Remarkably, if the surface scattering contribution is weak, then it
is possible to map the conduction problem of a long two-dimensional
(2D) rough waveguide to that of a 1D Anderson model of localization,
with the disorder potential determined by the surface roughness
\cite{firsttime,freilikher}.  Initially this mapping was established
for one-mode scattering but later it was generalized for any number
of modes in the transverse direction \cite{izoptic}.  As such, a
quantum wire with white-noise surface disorder will have zero
conductance if the localization length is much smaller than the wire
length. However, in reality the surface disorder of a rough quantum
wire always contains correlations. As a result it becomes
interesting and necessary to understand the conductance properties
in rough quantum wires with their surface disorder modeled by
colored noise. This has motivated several pioneering theoretical
studies \cite{firsttime,freilikher,izoptic,izprl,Alberto,Bagci}.
Under certain approximations the theoretical studies predicted
localization-delocalization transitions of electrons in 2D
waveguides with colored surface disorder. Some theoretical details
were tested by examining the eigenstates of a closed system with
rough boundaries \cite{iznum}. Moreover, the predicted mobility edge
due to colored disorder was recently confirmed in a microwave
experiment \cite{Kuhl}.

Using a reaction matrix formalism for direct scattering
calculations, here we computationally study the conductance
properties of rough quantum wires with colored surface disorder. The
motivation is threefold. First, though the dependence of the
localization length upon the correlation function of surface
roughness is now available from theory, how the more measurable
quantity, namely, the conductance of the waveguide, depends on
colored surface roughness has not been directly examined. This issue
can be quite complicated when the localization length becomes
comparable to the waveguide length.  Second, computationally
speaking it is possible to consider any kind of colored surface
disorder, thus realizing interesting circumstances that are not
readily testable in today's experiments. Indeed, in our
computational study we can create rather arbitrary structures in the
surface disorder correlation function and then examine the
associated conductance properties. Third, direct computational
studies allow us to predict some interesting conductance properties
that have not been treated theoretically or go beyond the validity
regime of available theories
\cite{firsttime,freilikher,izoptic,izprl,izprb}. For example, we
shall study the conductance properties for very strong surface
roughness, for rough bended waveguides, and for scattering energies
that are close to a shifted threshold value for transmission.  The
long-term goal of our computational efforts would be to explore the
usefulness of colored surface disorder in controlling the
conductance properties.

%We show the range of parameters of our model where one can get
%transition from localized to delocalized states or vice versa. We
%also look at the transport properties of electrons outside the range
%of assumptions used in predictions. Specifically we look at strong
%roughness, roughness with mean different than zero and energy
%regimes where inverse electron wavevector is larger than length of
%the wire. We point out the possibility of selective transmission or
%localization for a given energy of incoming electrons.

%The ability of controlling transport through long wires has immediate application on some
%of the important problems in literature. We will comment on heat conduction through nanowires,
%spin accumulation and Bose-einstein  condensates on optical lattices.

This paper is organized as follows. In Sec. II we describe the
scattering model of a quantum wire with colored surface disorder.
Therein we shall also briefly introduce the methodology we adopt for
the scattering calculations.  In Sec. III we present detailed
conductance results in a variety of one-mode scattering cases and
discuss these results in connection with theory.  Concluding remarks
are made in Sec. IV.

\section{Quantum scattering in waveguides with colored surface disorder}
\subsection{Modeling waveguides with colored surface disorder}

\begin{figure}
\includegraphics[width=10cm]{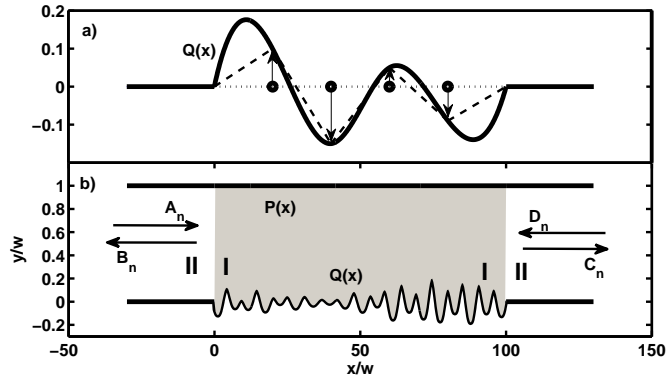}
\caption{Schematic plot of a 2D rough waveguide that models rough
quantum wires.  (a) The generation of a rough surface is illustrated
using $M=4$ random shifts in the transverse direction. (b) One
waveguide geometry with a straight upper boundary $y=P(x)=1$ and a
rough lower boundary $y=Q(x)$.  Scattering occurs in region I (gray
area) and region II denotes the left and right leads. Arrows
indicate the direction of incoming and outgoing electron waves.
$A_{n}$, $B_{n}$, $C_n$ and $D_{n}$ are quantum amplitudes (see Eq.
(8)).}
\end{figure}

We treat quantum wires as a long 2D waveguide as illustrated in Fig.
1(b). The scattering coordinate is denoted $x$ and the transverse
coordinate is denoted $y$. The width of the waveguide is denoted $w$
and the length is denoted $L$.   In all the calculations $L=100w$
and $w$ is set to be unity. That is, we scale all length by the
waveguide width.   The upper and lower boundaries of the waveguide
are described by $y=P(x)$ and $y=Q(x)$.  The case in Fig. 1(b)
represents a situation where the upper boundary is a straight line
($P(x)=1$) and the lower boundary is rough.  As in our other studies
of rough waveguides \cite{akgucprb06,akgucprb08}, we form a rough
waveguide boundary in three steps. First, we divide a rectangular
waveguide into $M$ pieces of equal length $L/M$. Second, the end of
each piece is shifted in $y$ randomly, with the random
$y$-displacement, denoted $\eta$, satisfying a Gaussian
distribution. Third, we use spline interpolation to combine those
sharp edges to generate a smooth curve $\eta(x)$ for either the
upper or the lower waveguide boundary. For the sake of clarity, Fig.
1(a) depicts this procedure with the number of random shifts being
as small as $M=4$. In all our calculations below we set $M=100$. In
Fig.~2(a) we show one realization of the surface roughness function
$\eta(x)$.

The function $\eta(x)$ may be characterized by its ensemble-averaged
mean $\overline{\eta}$ and its self-correlation function
$C_{\eta}(x-x')$, i.e.,
\begin{eqnarray}
\overline{\eta}=\langle\eta(x)\rangle&=&0, \nonumber \\
\langle\eta(x)\eta(x')\rangle&=&\sigma^2 C_{\eta}(x-x'),
\end{eqnarray}
where $\sigma$ is the variance of $\eta(x)$. In the limit of white
noise roughness, $C_{\eta}(x-x')$ is proportional to $\delta(x-x')$.
But more typically, $C_{\eta}(x-x')$ decays at a characteristic
length scale, called the correlation radius $R_c$.  For our case
here, because the randomness is introduced after dividing the
waveguide into $M=100$ pieces, the correlation length $R_c$ of the
$\eta(x)$ we construct here is of the order of $L/M \sim w$. This
length scale is comparable to the wavelength of the scattering
electrons in the one-mode regime.

One tends to characterize the strength of the surface roughness by
the variance $\sigma$ defined above.  However, in practice it is
better to use the maximal absolute value of $\eta(x)$, denoted
$|\eta_{\text{max}}|$, to characterize the roughness strength. This
is because for strong roughness with a given variance, there is a
possibility that some of the random displacements become too large
such that the waveguide may be completely blocked.  Recognizing this
issue, we first set a value of $|\eta_{\text{max}}|$ and then, after
having generated a roughness function $\eta(x)$ based on spline
interpolation, rescale $\eta(x)$ such that its maximal absolute
value is given by $|\eta_{\text{max}}|$ .

The roughness function $\eta(x)$ obtained above does not have any
peculiar features.  There are a number of ways to introduce some
structure to the correlation function $C_\eta(x-x')$.   In Ref.
~\cite{generator} a filtering function method was proposed to
produce a power-law decay of $C_\eta(x-x')$ from white noise. Here
we adopt the approach {\bf used in Ref. \cite{oldexp},} which  is based
on the convolution theorem of Fourier transformations. In
particular, the discrete form of autocorrelation function of
$\eta(x)$ is defined as
\begin{equation}
C'_\eta(\frac{mL}{N})=\sum_{n=1}^{N-m-1}\eta\left(\frac{(n+m)L}{N}\right)\eta\left(\frac{nL}{N}\right)
\label{chim}
\end{equation}
where $m=-N+1,\cdots,-1,0,1,\cdots,N-1$, $N$ is the total number of
grid points along $x$, and $c$ is a normalization constant such that
$C_{\eta}(0)=1$ \cite{corr-note}. In Fig.~2(b) we show the
autocorrelation function for the surface roughness function depicted
in Fig.~2(a). The autocorrelation drops from its peak value to near
zero at a scale of $R_c\sim 0.7 w$, which is much smaller than the
waveguide length.
%Note that due to our choices $L=100w$, $M=100$
%and $w=1$, Eq. (\ref{chim}) reduces to
%\begin{equation}
%\chi(m)=c\sum_{n=1}^{M-m-1}\eta(n+m)\eta(n).
%\end{equation}

\begin{figure}
\includegraphics[width=18cm]{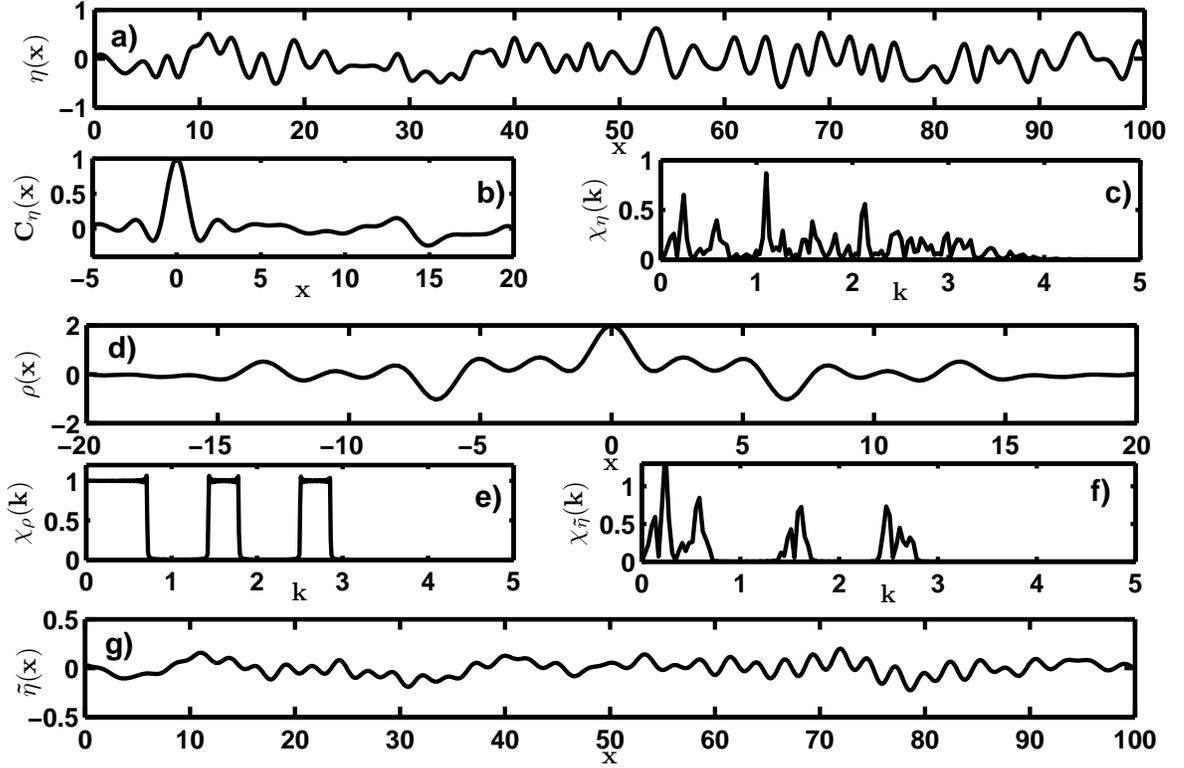}
\caption{(a) One realization of the surface roughness function
$\eta(x)$, with the method described in detail in the text. (b) The
associated autocorrelation function $C_{\eta}({x})$. (c) Surface
structure factor $\chi_{\eta}(k)$ obtained from the $C_{\eta}(x)$
shown in panel (b). (d) A function $\rho(x)$ that will be used to
introduce additional correlations via convolution.  (e) The Fourier
transform of the $\rho(x)$ shown in panel (d).  (f) The structure
factor $\chi_{\tilde{\eta}}(k)$ obtained from a convolution between
$\rho(x)$ and $\eta(x)$.  (g) The new surface roughness function
$\tilde{\eta}(x)$, with correlations that are absent in $\eta(x)$.}
\end{figure}

As will be made clear in what follows, it is important to consider
the Fourier transform of $C_{\eta}(x)$, i.e., the autocorrelation
function in the Fourier space.  This important quantity is denoted
$\chi_{\eta}(k)$, where $k$ is the wavevector conjugate to $x$.
Using the Fast Fourier transform of $C_\eta(x)$, $\chi_{\eta}(k)$
can be evaluated as follows:
\begin{equation}
\chi_{\eta}(k)=\sum_{j=1}^{2N}C_\eta(\frac{jL}{2N})\exp\left[\frac{-i2\pi
(j-1)(m-1)}{2N}\right],
\end{equation}
where the value of $k$ on the left side is determined by the value
of $m$ on the right side via $k=(2(m/2N)-1) (2 \pi N/2L)$. In our
calculations we choose $N=1024$. Note that $\chi_{\eta}(k)$ is a
real function due to the evenness of $C_{\eta}(x)$.  The real
function $\chi_{\eta}(k)$ is called below the structure factor of
the surface roughness. Figure 2(c) shows the structure factor
$\chi_{\eta}(k)$ obtained from the correlation function shown in
Fig. 2(b).
% and  $k=\frac{2\pi q}{4LN}$ is the wavevector defined in
%the interval $-\pi/L<k<\pi/L$.

Additional  correlations in the surface disorder can now be
generated by modulating the structure factor $\chi_{\eta}(k)$.
Because the structure factor $\chi_{\eta}(k)$ for a single
realization is equivalent to the square of the Fourier transform of
$\eta(x)$, we may imprint interesting structures onto
$\chi_{\eta}(k)$ by convoluting $\eta(x)$ with some filtering
function.  Consider then the function $\rho'(x)=\sin(ax)/ax$ with
$a>0$. Its Fourier transform is a step function of $|k|$,
\cite{izoptic} with a height $\pi/a$ and the step edge located at
$|k|=a$.  Consider then a combination of $n$ such functions, i.e.,
\begin{equation} \rho(x)=\sum_n A_n\frac{\sin(|a^r_n|
x)-\sin(|a^l_n| x)}{|a^r_n|  x},
\end{equation} where $A_n$, $a^r_n$, $a^l_n$ are predefined
parameters. Then the Fourier transform
 will be $\pi/A_n$ if $a_n^r>|k|> a_n^l$ or $a_n^r<|k|< a_n^l$; and
zero otherwise.  If we now consider the following roughness function
\cite{maxnote},
\begin{equation}
\tilde{\eta}\left(\frac{mL}{N}\right)=\sum_{n}
\rho\left(\frac{nL}{N}\right)\eta\left(\frac{(m-n)L}{N}\right),
\end{equation}
then according to the convolution theorem, we have
\begin{eqnarray}
|\chi_{\tilde{\eta}}(k)|^{1/2} \sim |\chi_{\rho}(k)|
|\chi_{\eta}(k)|^{1/2}, \label{chitilde}
\end{eqnarray}
where $\chi_{\tilde{\eta}}(k)$ is the structure factor for the new
surface roughness function $\tilde{\eta}(x)$. As such, the structure
of $\chi_{\rho}(k)$ is directly imprinted on
$\chi_{\tilde{\eta}}(k)$. That is, computationally speaking,
arbitrary modulation can be imposed on the structure factor by
filtering out the unwanted components and magnifying other desired
structure components.   Below we apply this simple technique to
create different kinds of surface roughness correlation windows and
then examine the conductance properties.  In Fig. 2(d) we show one
example of $\rho(x)$. Its Fourier transform amplitude, as shown in
Fig. 2(e), displays two windows.  As shown in Fig. 2(f), this
double-window structure is passed to $\chi_{\tilde{\eta}}(k)$ due to
Eq. (\ref{chitilde}). Finally, in Fig. 2(g) we show the surface
roughness function $\tilde{\eta}(x)$, which obviously contains more
correlations than the old surface roughness function $\eta(x)$ shown
in Fig. 2(a).

\subsection{Reaction matrix and scattering matrix }
Here we briefly describe how we calculate the electron conductance
of a rough 2D waveguide as described above.  The Hamiltonian for the
quantum transport problem  is given by
    \begin{equation}
    \widehat{H}=-\frac{\hbar^2}{2m^*}\left(\frac{\partial^2}{\partial x^2}+\frac{\partial^2}{\partial y^2}\right)
    +V_c(x,y),
    \label{eq:se}
\end{equation}
where $m^*$ is the electron effective mass and $V_c(x,y)$ represents
a hard wall confinement potential. That is, $V_c(x,y)$ is zero in
$Q(x)<y<P(x)$ and becomes infinite otherwise.  In our early work
\cite{akgucprb06,akgucprb08} we formulated such a waveguide
scattering problem in detail in terms of the so-called reaction
matrix method.
%use in following calculations. %We use atomic units such that
%$\hbar=2m*=1$ and length units are in terms of channel width $w$.
In the reaction matrix method we first expand scattering state in
the scattering region (region I, gray area in Fig. 1(b)) in terms of
a complete set of basis states.  The basis states are obtained by
transforming the rough waveguide into a rectangular one, with the
expense of a transformed Hamiltonian with extra surface dependent
terms. The solutions in the leads (region II, Fig. 1(b)) are given
by
\begin{eqnarray}
\Psi_{\emph{l}} = \left (A_n \frac{e^{ik_n x}}{\sqrt{k_n}}
-B_n\frac{e^{-ik_n x}}{\sqrt{k_n}}\right)
\sin(\frac{n\pi y}{w}); \nonumber \\
\Psi_{\emph{r}} = \left (C_n \frac{e^{ik_n x}}{\sqrt{k_n}}
-D_n\frac{e^{-ik_n x}}{\sqrt{k_n}}\right) \sin(\frac{n\pi y}{w})
\label{sca}
\end{eqnarray}
for the left and right leads, respectively. Here $n$ is the index
for the modes in the transverse direction, and the wavevector $k_n$
is given by
\begin{equation}
k_n=\sqrt{\frac{2mE}{\hbar^2}-\left(\frac{n\pi}{w}\right)^2},
\label{kf}
\end{equation}
where $E$ is the initial electron energy. The scattering
coefficients $A_n$, $B_n$, $C_n$ and $D_n$ in Eq. (\ref{sca}) are
determined by the scattering matrix $S$, which relates the outgoing
states to the incoming states. Specifically,
\begin{equation}
\left(\begin{array}{c} B_n
\\C_n\end{array}\right)=\left(\begin{array}{cc} r& t \\t'&
r'\end{array}\right) \left(\begin{array}{c}A_n
\\D_n\end{array}\right),
\end{equation}
where the submatrices $r$ and $r'$ denote the reflection matrix and
$t$ and $t'$ denote the transmission matrix.  In the case of
one-mode scattering ($n=1$) considered below, $k_{1}$ will be simply
denoted as $k$, with $0<kw/\pi<\sqrt{3}$. The $S$ matrix is related
to the so-called $R$-matrix in the reaction matrix method as
follows,
\begin{equation}
S=\frac{I_{2m} -i K R K}{ I_{2m} + i K R K}.
 \end{equation}
where $m$ is maximal number of propagating modes and  $I_{2m}$ is a
$2m\times 2m$ unit matrix, and $K$ is $2m\times 2m$ diagonal matrix
with diagonal elements determined by the wavevector associated with
each scattering channel \cite{akgucprb06,akgucprb08}. Once the $S$
matrix is obtained from the $R$ matrix, the conductance is
calculated by $G=G_0 \text{Trace}(tt')$, where $G_0=e^2/(2\hbar)$ is
the conductance quanta.  Note that in our calculations  we include
about 10 evanescent modes
 %Since we use a long waveguide geometry,
%basis states in transformed space should be arranged accordingly.
though we focus on the energy regime where only one mode in the
$y$-direction admits propagation along $x$.  As to the number of
basis states we use in describing the transformed rectangular
waveguide, we use 1000 basis states for the $x$ degree of freedom
and $4$ basis states for the $y$ degree of freedom.  Such a large
number of basis states is for a good description of the scattering
wave function inside the waveguide, and this number should not be
confused with the number of propagating modes or evanescent. Good
convergence is obtained in our calculations.  Note also that due to
the large number of basis states used in the scattering direction,
the Fourier transform techniques developed in Ref. \cite{akgucprb06}
is especially helpful.

\section{Effects of Colored Surface Disorder On Conductance}

%% put the variance of eta(x) back into the previous section...)
%% then we do the rest...
With the mapping between the scattering problem in 2D waveguide and
1D Anderson's model \cite{firsttime,freilikher},
%The
%transformation of surface disorder into 1D Anderson model enables us
%to assign a localization length for the 2d waveguide. The surface
%function $\eta$ is characterized as follows:
%\begin{equation}
%<\eta(x)>=0, \  \  <\eta(x)\eta(x')>=\sigma^2 \chi_{\eta}(x-x').
%\end{equation}
%where $\sigma^2=<\eta^2>-<\eta>^2$ is the strength of disorder,
%correlation function decays with a correlation radius,
%$R_c$ and averaging
%is done over an ensemble of surfaces.  Surface function $\eta$ is regarded as
% the Gaussian white noise if it has a delta function autocorrelation, i.e.
% $\chi_\eta(|x-x'|)=R_c\delta(x-x')$ with surface structure factor having contribution
% from all the wavevectors. In numerical calculations here we have  $R_c=1.4$ and
% higher wavevector contributions are non existent.
early theoretical work \cite{firsttime,izoptic} established that the
localization length $L_{\text{loc}}$ of the 2D waveguide problem is
given by
\begin{equation}
 L_{\text{loc}}^{-1}=\frac{\sigma^2
 \pi^4}{w^6}\frac{\chi(2k)}{(2k)^2},
 \label{eq:loc}
\end{equation}
where $\chi(2k)$ is either the structure factor $\chi_{\eta}(2k)$ or
the new structure factor $\chi_{\tilde{\eta}}(2k)$ after a
convolution procedure.
 If $L_{\text{loc}}>L$, a transmitting state is expected and if $L_{\text{loc}}<<L$, then
the electron can only make an exponentially small contribution to
the conductance. As such, one expects transmitting states when the
structure factor $\chi(2k)$ is essentially zero; and negligible
conductance if $\chi(2k)$ is significant and if $\sigma$ is not too
small. This suggests that the conductance properties can be
manipulated by realizing different surface roughness functions.
%In
%addition, as is also clear from Eq. (\ref{eq:loc}), because by
%construction the scale of surface roughness in our rough waveguide
%model is bounded by $R_c$, conductance properties of high energy
%electrons (opening many modes) will not be much affected by the
%surface roughness modeled here.

Equation (\ref{eq:loc}) is obtained under a weak electron scattering
approximation (Born approximation). As such, the theoretical result
of Eq. (\ref{eq:loc}) may not be valid if $\sigma$ is not small as
compared with $w$ or if the scattering electron is close to the
threshold value of channel opening. Another assumption in the theory
is that $L_{\text{loc}}$ should be much greater than $R_{c}$, the
radius of the surface correlation function $C_{\eta}(x)$.

\subsection{Straight Rough Waveguides}
%For
%one-mode scattering this condition is equivalent to $k_F\sigma<<1$.
However, in our computational studies we will examine some
interesting cases that are evidently beyond the validity regime of
the theory. For example, the strength of the surface disorder may
not be small and the scattering energy may be placed in the vicinity
of a shifted channel opening energy. %As explained earlier, this also
%motivated us to use the maximal random $y$-shift of $\eta(x)$,
%instead of $\sigma$, to characterize the strength of surface
%roughness.
% T real
%values and evanescent modes for imaginary values.
%In following
%calculation we have have one mode propagation others evanescent
%which restricts propagating wavevector as $0<k_Fw/\pi<\sqrt{3}$.
%Therefore the assumption of $\sigma<<w$ is same as $k_F\sigma<<1$.
%{\bf Gursoy explain the reason of the following statement:}. That is
%the comparison near channel opening is outside the analytical
%predictions.
%Another assumption in the theory is that
%$L_{\text{loc}}$ should be much greater than $R_{c}$, the radius of
%the surface correlation function $C(x)$.
%This means we cant
%manipulate higher energies in a mode since that requires random
%surface is high grained which cant be resolved by the electron wave.
%{\bf Gursoy: why do we need the following sentence? Is the following
%relevant to our results below?} The two assumptions can be
%summarized by the relation $
%\max(k_F^{-1},R_c)<<\min(L_{\text{loc}},L)$ \cite{notes}.
\begin{figure}
\includegraphics[width=16cm]{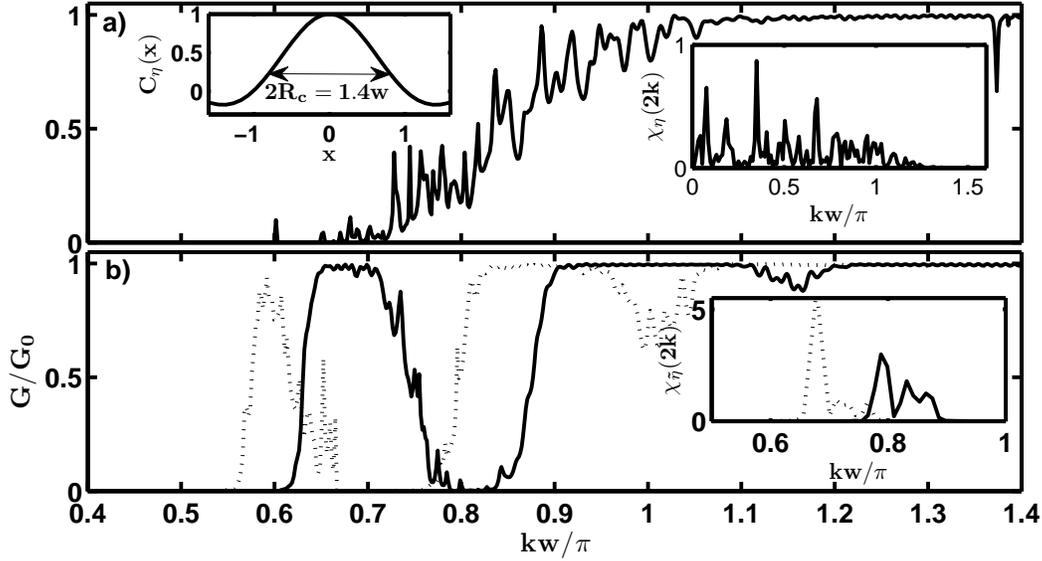}
\caption{(a) Conductance of rough waveguides vs $k=k_{1}$ (see Eq.
(9)). The upper boundary is flat, i.e., $P(x)=1$, and the lower
boundary is given by $Q(x)=\eta(x)$, with a surface disorder
strength characterized by $|\eta_{\text{max}}|=0.2w$ and the
variance $\sigma$ given by $\sigma=0.0779w$, $\sigma=0.0802w$ and
$\sigma=0.0773w$.
 Inset on the right shows the structure factor of the surface
roughness in one single realization.  (b) Same as in panel (a), but
two cases with different surface structure factor obtained from a
convolution approach are plotted, using solid and dashed lines. All
conductance curves here are averaged over three realizations.}
\end{figure}

In Fig. 3(a) we show conductance results averaged over three
realizations of a rough waveguide, with a flat upper boundary
$P(x)=1$ and a rough lower boundary $Q(x)=\eta(x)$. The strength of
the surface disorder is characterized by $|\eta_{\text{max}}|=0.2w$.
Due to our procedure in generating a fixed
$|\eta_{\text{max}}|=0.2w$, the variance $\sigma$ of the surface
function in each single realization of the surface function will
change slightly. For the three realizations used in Fig. 3(a),
$\sigma=0.0779w$, $0.0802w$ and $0.0773w$.
%The strength of
%the surface disorder is given by $\sigma=0.0779w$ which is chosen
%based on a pre-set value of $|\eta_{\text{max}}|=0.2w$.
As is clear
from Fig. 3(a), there exists a threshold $k\sim 0.6 \pi/w$ beyond
which the system becomes transmitting (this threshold will be
explained below). In the transmission regime the conductance shows a
systematic trend of increase as the wavevector $k$ increases. The
inset of Fig. 3(a) shows $\chi_{\eta}(2k)$, one key term in Eq.
(\ref{eq:loc}). The characteristic magnitude of $\chi_{\eta}(2k)$
for the shown regime of $k$ is $\sim 0.3$.  Using Eq.
(\ref{eq:loc}), one obtains that the localization length
$L_{\text{loc}}$ is comparable to $L=100w$. This prediction is hence
consistent with our computational results that demonstrate
considerable transmission.

Next we exploit the convolution technique described above to form
new rough surfaces described by $\tilde{\eta}(x)$.  In particular,
the inset of Fig. 3(b) shows two sample cases with distinctively
different surface structure factors. In one case (dotted line)
$\chi_{\tilde{\eta}}(2k)$ has significant values in the interval
$0.67<kw/\pi<0.8$. Indeed, during that regime the value of
$\chi_{\tilde{\eta}}(2k)$ is many times larger than the mean value
of $\chi_{\eta}(2k)$ in the case of Fig. 3(a). In the other case
(solid line) $\chi_{\tilde{\eta}}(2k)$ is  large only in the regime
of $0.75<kw/\pi<0.9$.  For these regimes, the theory predicts that
the localization length to be much smaller than the waveguide length
and hence vanishing conductance.   This is indeed what we observe in
our computational study.  As shown in Fig. 3(b), either the dotted
or the solid conductance curve display a sharp dip in a regime that
matches the main profile of $\chi_{\tilde{\eta}}(2k)$.

In addition, similar to what is observed in Fig. 3(a),  Fig. 3(b)
also displays a transmission threshold. Take the dotted line in Fig.
3(b) as an example.  For $kw/\pi<0.55$, there is no transmission at
all, even though $\chi_{\tilde{\eta}}(2k)$ in that regime is
essentially zero. This suggests that this threshold behavior is
unrelated to surface roughness details. Rather, it can be considered
as a non-perturbative result that is not captured by Eq.
(\ref{eq:loc}). To qualitatively explain the observed threshold, we
realize that due to the relatively strong surface roughness, the
effective width of the waveguide decreases and as a result, the
effective mode opening energy increases \cite{akgucprb06}.  For
$|\eta_{\text{max}}|=0.2w$, we estimate that the effective width of
the waveguide is given by $w_{\text{eff}}=
w-|\eta_{\text{max}}|=0.8w$. Hence, the corrected mode opening
energy $E$ is now given by $(\hbar^2/2m^{*})(\pi/0.8w)^2$. Using Eq.
(\ref{kf}), this estimate gives that, regardless of the surface
roughness details, the threshold $k$ value for transmission is $\sim
0.75\pi/w$, which is close to what is observed in Fig.~3. Such an
explanation is further confirmed below. This also demonstrates that
the maximal value of $|\eta(x)|$ is an important quantity to
characterize the surface roughness strength.  Of course, the exact
dependence of the effective waveguide width upon $\eta_{max}$ is
beyond the scope of this work \cite{kro}.

The results in Fig. 3 show that even when the surface roughness is
strong enough to significantly shift the threshold energy for
transmission, the surface structure factor may still be well
imprinted on the conductance curve. Moreover, the resultant windows
of the conductance curves in Fig. 3 are seen to match the location
of the structure factor peak.  Nevertheless, one wonders how such an
agreement might change if we tune the strength of the surface
roughness.

\begin{figure}
\includegraphics[width=16cm]{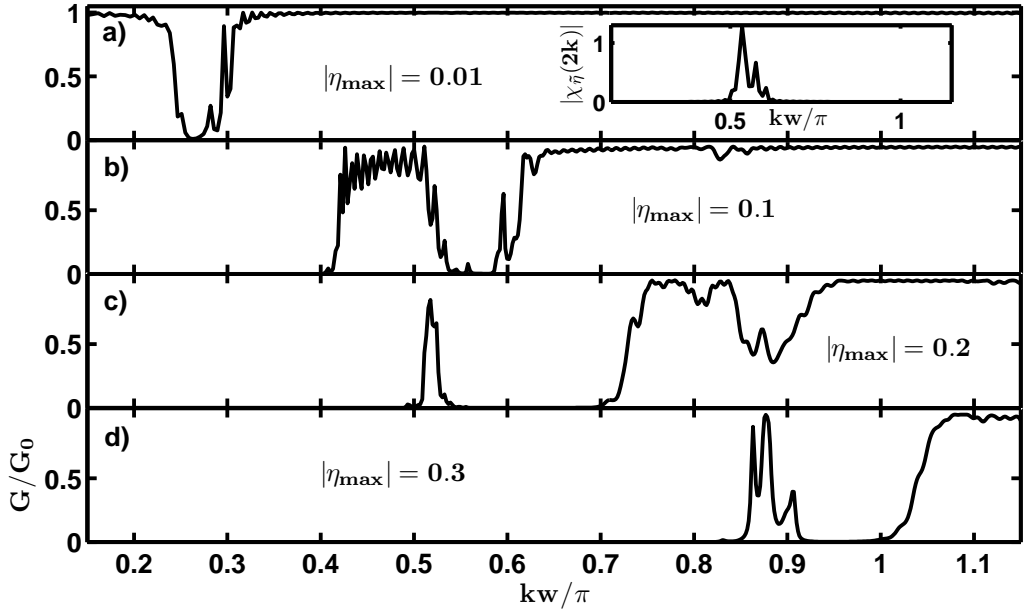}
\caption{Conductance in  rough waveguides with increasing strength
of surface roughness. The waveguide geometry is similar to what is
considered in Fig. 3, but with a different structure factor.
$|\eta_{\text{max}}|=0.01$ and $\sigma=0.0046w$ in (a), $|\eta_{\text{max}}|=0.1$ and $\sigma=0.0400w$ in (b),
$|\eta_{\text{max}}|=0.2$ and $\sigma=0.0779w$ in (c), and $|\eta_{\text{max}}|=0.3$ and $\sigma=0.1255w$ in
(d). Inset shows the surface structure factor used in (a)--(d).}
\end{figure}

To that end we examine in Fig. 4  four scattering cases with
increasing roughness strength, with $|\eta_{\text{max}}|=0.01w$ and
$\sigma=0.0046w$ in Fig. 4(a) (representing a case with quite weak
surface roughness), $|\eta_{\text{max}}|=0.1w$ and $\sigma=0.0400w$
in Fig. 4(b), $|\eta_{\text{max}}|=0.2w$ and $\sigma=0.0779w$ in
Fig. 4(c), and $|\eta_{\text{max}}|=0.3w$ and $\sigma=0.1255w$ in
Fig. 4(d) (representing a case with very strong surface roughness).
%In each case, we study the conductance for two types of roughness
%structure factor, with peak of the first $\chi_{\tilde{\eta}}(k)$
%located at higher $k$ values and the peak of the other located at
%quite small $k$ values.
The main profile of the structure factor is also shifted closer to
the threshold regime observed in Fig. 3.  For the case with
$|\eta_{\text{max}}|=0.3w$, the theory based on the weak roughness
assumption is not expected to hold. Indeed, In Fig. 4(d) the
transmission threshold is right shifted further to the high energy
regime as compared with those seen in Fig. 3 or other panels in Fig.
4. Nevertheless, we still observe a clear window of almost zero
conductance, but now with its location also significantly shifted as
compared with the profile of $\chi_{\tilde{\eta}}(2k)$.  For the
case of $|\eta_{\text{max}}|=0.2w$ in Fig. 4(c), it is somewhat
similar to the dotted line in Fig. 3(b), consistent with the fact
that they have the same roughness strength. However, because here
the location of the peak of $\chi_{\tilde{\eta}}(2k)$ is close to
the threshold $k$ value, the zero conductance window is also near
this threshold:  the conductance curve rises when $k$ exceeds the
threshold and then it quickly drops to zero again.   For the case of
$|\eta_{\text{max}}|=0.1w$, its zero conductance window shown in
Fig. 4(b) is narrower than those in Fig. 4(c) and Fig. 4(d),
consistent with our intuition. Somewhat surprising is the case shown
in Fig. 4(a), where the roughness strength is weak and the energy
threshold for transmission is almost unaffected. But still, a narrow
window for very small conductance is clearly seen in Fig. 4(a). This
result is unexpected, because if one applies Eq. (\ref{eq:loc})
directly, one would predict that no such conductance window should
occur for $|\eta_{\text{max}}|=0.01$. Further,  the conductance
window in Fig. 4(a) is much left-shifted as compared with the
profile of $\chi_{\tilde{\eta}}(2k)$ (inset of Fig. 4(a)). Similar
results are obtained in other realizations of the surface roughness
function $\eta(x)$ that have similar profile of the structure factor
\cite{smallknote}. Such a remarkable deviation from the theory, we
believe, is due to a breakdown of the Born approximation in deriving
Eq. (\ref{eq:loc}). Indeed, the conductance window for
$|\eta_{\text{max}}|=0.01$ is located in a regime of very low
scattering energy, and is hence not describable by a theory based on
the Born approximation. Certainly, it should be of considerable
interest to experimentally study the conductance windows in these
cases of weak surface roughness.

To further confirm that the conductance windows observed here are
due to the colored surface disorder, we note that if we consider a
surface function as that shown in the inset of Fig. 3(a), then all
the conductance windows shown here indeed disappear and the results
will become similar to what is shown in Fig. 3(a).
%This shift is not linear and it breaks for the stronger
%surface disorder which makes it not a flexible way to control
%transmission.
\subsection{Bended Rough Waveguides}
\begin{figure}
\includegraphics[width=16cm]{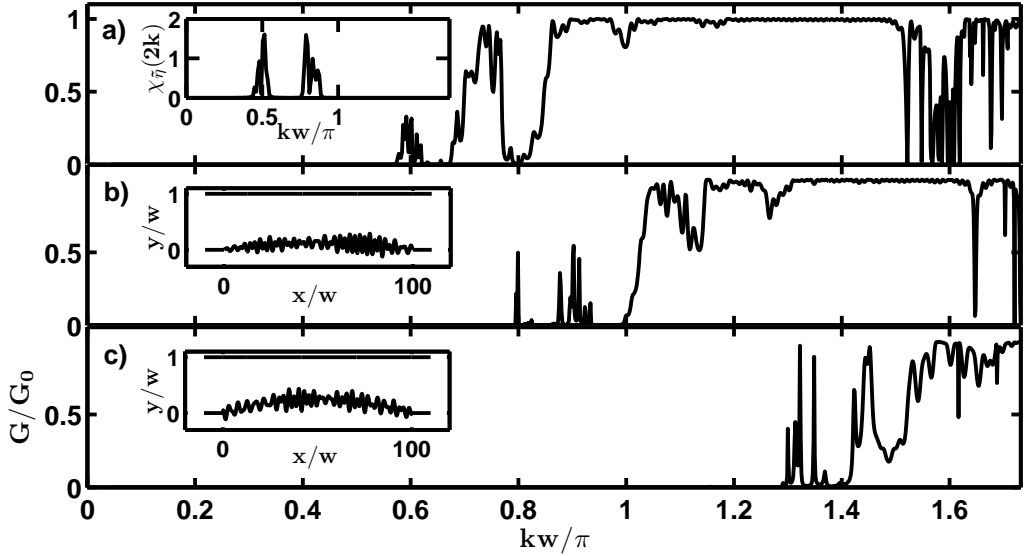}
\caption{Conductance properties of rough bended waveguides, with a
straight upper boundary $P(x)=1$ and a bended lower boundary given
by $Q(x)=4a(L/2-y)^2/L^2+\tilde{\eta}(x)$.  The waveguide curvature
parameter $a$ is given by $a=0$ in (a),
 $a=0.5$ in (b), and $a=1.0$ in (c).  The inset in
 (a) shows the surface structure factor used in all the calculations, and the insets in  (b) and (c)
 show the associated waveguide geometry.}
\end{figure}

In Fig. 5 we examine the conductance properties of a bended rough
waveguide (Fig. 5(b) and Fig. 5(c)) as compared with those of a
straight rough waveguide (Fig. 5(a)). In all the three cases shown
in Fig. 5, the upper boundary is given by $P(x)=1$;  and the lower
boundary is a parabolic curve plus random fluctuations, i.e.,
$Q(x)=4a(x-L/2)^2/L^2+\tilde{\eta}(x)$; with $a=0$ in Fig. 5(a),
$a=0.5$ in Fig. 5(b), and $a=1.0$ in Fig. 5(c).  As to the structure
factor of $\tilde{\eta}(x)$, it is assumed to be of a double-window
form as shown in the inset of Fig. 5(a), with $|\eta_{\text{max}}|$
the same as in Fig. 3. In the case of a straight rough waveguide,
this double-window structure factor creates an analogous
double-window structure in the conductance curve (Fig. 5(a)), with
its location matching the profile of the structure factor.
Interestingly, as we introduce a curvature in the lower boundary in
Fig. 5(b), the double-window structure survives but shifts
considerably toward higher $k$ values. In Fig. 5(c), the curvature
of the rough waveguide further increases, the transmission threshold
value of $k$ also increases (as expected), and the fingerprints of
the double-window structure factor can still be seen in the
conductance curve.  We have also checked if we create three windows
in the structure factor, then three windows in the conductance
curves can be induced as well, with their locations controllable by
tuning the curvature of the bended waveguide.

\begin{figure}
\includegraphics[width=16cm]{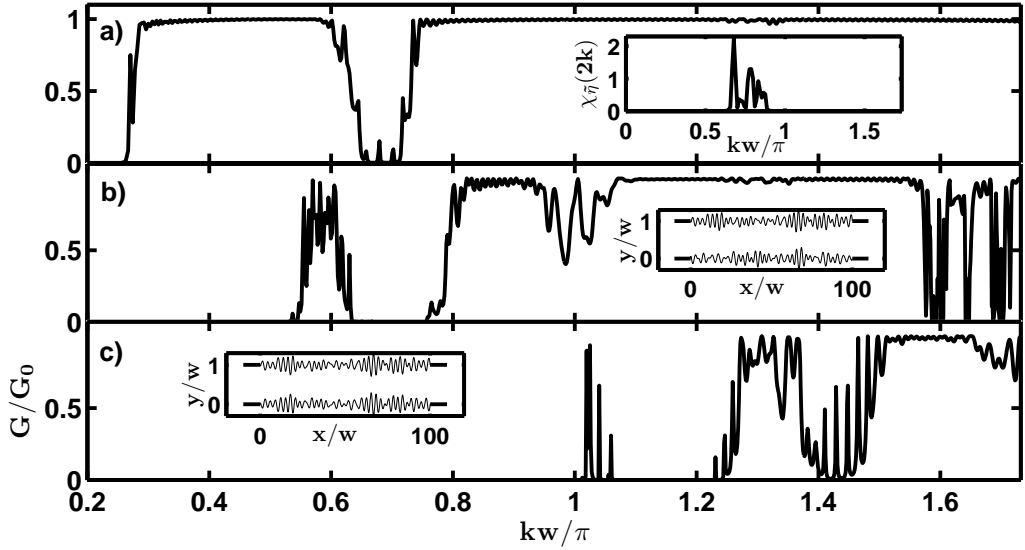}
\caption{ (a) Conductance of a rough waveguide with symmetric
surfaces, modeled by an upper boundary $P(x)=1+\tilde{\eta}$ and a
lower boundary $Q(x)=\tilde{\eta}(x)$. Inset shows the surface
structure factor used in all the calculations here.  (b) Conductance
of a rough waveguide with
 two uncorrelated surfaces, modeled by an upper boundary $P(x)=1+\tilde{\eta}'(x)$ and
 a lower boundary of $Q(x)=\tilde{\eta}(x)$. Inset shows the associated waveguide
 geometry.
 (c) Conductance of a
rough waveguide with two antisymmetric surfaces, modeled by an upper
boundary $P(x)=1-\tilde{\eta}(x)$ and a lower boundary
$Q(x)=\tilde{\eta}(x)$. Inset shows the associated waveguide
geometry.}
\end{figure}
Finally we consider waveguides with both upper and lower boundaries
being rough.  Interestingly, in this case a more sophisticated
theory \cite{izprb} shows that the scattering can be regarded as the
scattering in a smooth waveguide plus an additional effective
potential. The theoretical electron mean free path, calculated using
a Green function averaged over different surfaces, is shown to be
contributed by different terms, due to different mechanisms called
amplitude scattering, gradient scattering, and square gradient
scattering \cite{izprb}. The importance of these terms depends on
whether the upper and lower boundaries are symmetric, uncorrelated
or anti-symmetric.  Motivated by this interesting prediction, here
we show in Fig. 6 three computational results, for symmetric (Fig.
6(a)), uncorrelated (Fig. 6(b)), and anti-symmetric boundaries (Fig.
6(c)), all with the same roughness strength as in Fig. 3.

%In
%particular, in Fig.~6(a) the waveguide boundaries are given by
%$P(x)=1+\eta(x)$ and $Q(x)=\eta(x)$. The surface structure factor is
%shown in the inset of Fig. 6(a), with the same roughness strength as
%in Fig. 3. In the case of Fig. 6(b), $P(x)=1+\eta'$ and $Q(x)=\eta$.
For the symmetric case,  the effective waveguide width is not
affected by the roughness. By contrast, for the antisymmetric case
of Fig. 3(c), the effective waveguide width is most affected. These
two simple observations explain why the threshold $k$ value for
transmission is the smallest in Fig. 6(a) and the largest in Fig.
6(c).  Even more noteworthy is how the structure factor of surface
roughness generates a conductance window for these three cases. In
particular, the window of the conductance drop in the symmetric case
(Fig. 6(a)) is narrower than that seen in Fig. 6(b) and Fig. 6(c).
%We believe that this is a manifestation of the fact that the
%localization length in the symmetric case has no contribution from
%the so-called gradient scattering mechanism \cite{izprb}.
Moreover, the conductance window in the anti-symmetric case is the
widest one, and is much shifted towards high $k$ values as compared
with the structure factor.  This large mismatch between the
conductance window and the peak location of
$\chi_{\tilde{\eta}}(2k)$  hence reflects clearly an effect from the
correlation between the two rough boundaries. Though our results
cannot be easily explained by the theoretical result of Eq.
(\ref{eq:loc}), they are consistent with the theoretical prediction
in Ref. \cite{izprb} that among the three cases of symmetric,
uncorrelated and anti-symmetric waveguides, the electron mean free
path in anti-symmetric waveguides should be the shortest.

%The standard theory of transport through a long surface disordered wire is given
%by DMPK equation. \cite{Mello}
%This theory gives the evolution of the probability density distribution
%of the scattering matrix when the length L of the system increases.
%The main assumption in DMPK equation is homogeneity of scattering matrix.
%At any position of waveguide phases of scattering matrix  eigenvalues are assumed
%to be distributed uniformly over all angles (fully chaotic). This assumption fails for a wire with colored surface disorder. We have checked the behavior of localized regions
%in conductance follow the DMPK predictions in our calculations. The crossover
% regions and transmitting (metalic) regions is not described correctly by DMPK theory.
\section{Discussion and Conclusion}
In this computational study we have focused on how the structure of
surface roughness impacts the conductance properties of electrons
propagating in a quantum wire modeled by a 2D waveguide. Our
conductance results are directly computed from a reaction matrix
approach.  An early theoretical result is hence confirmed by
detailed behavior of the conductance, a quantity that should be
measurable in experiments. In addition, our results for symmetric,
uncorrelated, and anti-symmetric rough waveguides are consistent
with a very recent theory \cite{izprb}.

Unlike in the bulk case, for quantum wires of limited length the
sensitive dependence of the localization length upon the structure
factor of surface roughness can be easily manifested in conductance
properties. Our direct scattering calculations show that this is
true, even for those interesting cases that are beyond the domain of
today's theory or have not been treated theoretically. We conclude
that conductance properties are easily controllable by engineering
the surface roughness of quantum wires.

Though we have focused on the transport behavior of electrons, we
believe that our methodology might be also useful for studies of
other types of wave propagation in disordered systems. In
particular, there is now a keen interest in understanding phonon
transport in rough quantum wires.
%Recent numerical studies \cite{Murphy, Baladin} observed
%anomalous heat transport in rough nanowires with small radius.
Recent computational work \cite{Murphy} and experiment work
\cite{Li} showed the importance of surface disorder in the heat
transport of thin silicon nanowires with a radius of $w=22$ nm. It
was also demonstrated experimentally that surface roughness can be
used to dramatically suppress heat conductivity \cite{thermo} and
hence enhance thermoelectric efficiency for thin silicon nanowires
with a radius about $w=50$ nm.  Our computational tools, together
with the guidance from the theory
\cite{firsttime,freilikher,izoptic,izprl,izprb}, might help answer
some important questions regarding to phonon transport in rough
nanowires.  Indeed, we conjecture that it should be possible to
design some colored surface to create conductance windows for
phonons, but not for electrons. If this is indeed realized, then
electron conductance is not much affected and phonon conductance
will be greatly reduced. This will be of vast importance to
thermoelectric applications.

Finally, we note that spin accumulation in quantum waveguides with
rough boundaries was recently studied in Ref. \cite{akgucprb08}.  It
should be interesting to see how colored surface disorder might have
some useful impact on spin accumulation effects or spin transport.

%Spin accumulation has been observed to occur at mode transitions
%which can be adjusted using correlated surface. Works in this area
%is underway. Finally recent experiments on bose-einstein systems
%\cite{bose, muller} indicates matter localization for a 1D optical
%lattices. It may be interesting to see the behavior of condensate
%with  optical lattice with colored noise disorder.

\acknowledgments We thank Prof. Li Baowen for interesting
discussions. This work was supported by the start-up fund (WBS grant
No. R-144-050-193-101/133), National University of Singapore, and
the NUS ``YIA" fund (WBS grant No. R-144-000-195-123), from the
office of Deputy President (Research \& Technology), National
University of Singapore. One of the authors, G.B. Akguc, thank F. M.
Izrailev for fruitful discussions. We also thank the Supercomputing
and Visualization Unit (SVU), the National University of Singapore
Computer Center, for use of their computer facilities.

%\appendix


\begin{thebibliography}{10}
\bibitem{corr-note} Note that to be more precise, one should compensate the
discretized correlation function defined in Eq. (2) by a factor of
$1/(N-m-1)$. However, this compensation has little effect here
because $N$, i.e., the total number of grid points, is large enough
and the correlation function itself decays fast (correlation radius
much smaller than the waveguide length).
\bibitem{maxnote} After the convolution, we will rescale the new
surface roughness function $\tilde{\eta}(x)$ such as its maximal
value is still given by $|\eta_{\text{max}}|$.
\bibitem{smallknote} Interestingly, for $|\eta_{\text{max}}|=0.01$,
if the location of the peak of $\chi_{\tilde{\eta}}(2k)$ is too
close to the channel opening energy, e.g., $\sim kw/\pi = 0.25$,
then it will be hard to identify a conductance window similar to
that shown in Fig. 4(a) as it will be buried in the channel opening
regime.
\end{thebibliography}
\end{document}